\documentclass[aps,prl,twocolumn,10pt,amsmath,amssymb,bibnotes,superscriptaddress,longbibliography]{revtex4-1}

\usepackage{graphicx, txfonts, pifont}
\usepackage{siunitx}
\usepackage[colorlinks,urlcolor=blue,citecolor=blue,linkcolor=blue]{hyperref}
\usepackage{tikz}

\newcommand{\tp}{\ensuremath{\tau_\varphi^{-1}}}
\newcommand{\ti}{\ensuremath{\tau_i^{-1}}}
\newcommand{\fg}{H\ensuremath{_2}/N\ensuremath{_2}~}
\newcommand{\tee}{\ensuremath{\tau_{ee}^{-1}}}
\newcommand{\txs}{\ensuremath{\tau_{xs}^{-1}}}

\begin{document}

\title{Activation of magnetic moments in CVD-grown graphene by annealing}

\author{Hyungki Shin}
\email{hshin@phas.ubc.ca}
	\affiliation{Stewart Blusson Quantum Matter Institute, University of British Columbia, Vancouver, British Columbia, V6T1Z4, Canada}
	\affiliation{Department of Physics and Astronomy, University of British Columbia, Vancouver, British Columbia, V6T1Z1, Canada}
\author{Ebrahim Sajadi}
	\affiliation{Stewart Blusson Quantum Matter Institute, University of British Columbia, Vancouver, British Columbia, V6T1Z4, Canada}
	\affiliation{Department of Physics and Astronomy, University of British Columbia, Vancouver, British Columbia, V6T1Z1, Canada}
\author{Ali Khademi}
	\affiliation{Stewart Blusson Quantum Matter Institute, University of British Columbia, Vancouver, British Columbia, V6T1Z4, Canada}
	\affiliation{Department of Physics and Astronomy, University of British Columbia, Vancouver, British Columbia, V6T1Z1, Canada}
	\affiliation{Present address: National Research Council Canada, Edmonton, Alberta T6G 2M9, Canada}

\author{Silvia L\"{u}scher}
	\affiliation{Stewart Blusson Quantum Matter Institute, University of British Columbia, Vancouver, British Columbia, V6T1Z4, Canada}
	\affiliation{Department of Physics and Astronomy, University of British Columbia, Vancouver, British Columbia, V6T1Z1, Canada}
\author{Joshua A. Folk}
\email{jfolk@physics.ubc.ca}
	\affiliation{Stewart Blusson Quantum Matter Institute, University of British Columbia, Vancouver, British Columbia, V6T1Z4, Canada}
	\affiliation{Department of Physics and Astronomy, University of British Columbia, Vancouver, British Columbia, V6T1Z1, Canada}
\date{\today}

\begin{abstract}

   Effects of annealing  on chemical vapor deposited graphene are investigated via a weak localization magnetoresistance measurement. Annealing at \SI{300}{\celsius} in inert gases, a common cleaning procedure for graphene devices \cite{lin2011graphene,cheng2011toward}, is found to raise the dephasing rate significantly above the rate from electron-electron interactions, which would otherwise be expected to dominate dephasing at 4 K and below \cite{PhysRevB.22.5142, kozikov2010electron, tikhonenko2008weak,tikhonenko2009transition, min2017asymmetric}.  This extra dephasing is apparently induced by local magnetic moments activated by the annealing process, and depends strongly on the backgate voltage applied.

\end{abstract}
 
\maketitle


 \section{I. Introduction}
 The first graphene samples made into electronic devices came from flakes exfoliated from bulk graphite, and were only a few microns in size \cite{neto2009electronic, geim2010rise}. The discovery that graphene could also be grown over large areas, using a chemical vapour deposition (CVD) process on metal films, opened up many technological possibilities that were unimaginable with exfoliated flakes. One  application proposed in the early days of graphene research was spintronics, making use of  potentially long spin lifetimes in carbon-based materials. Although the hope of graphene spintronics persists, experimental realizations remain less impressive than theoretical proposals \cite{RevModPhys.92.021003,han2011spin}.
 
 Despite the promise of CVD-grown graphene for technology, this growth technique tends to yield samples with more defects, such as carbon vacancies and domain boundaries, compared to exfoliated graphene \cite{lee2017review}. In fact, defects like those found in CVD graphene are believed to be a potential source of magnetic moments that would spoil spintronic applications\cite{yazyev2007defect}.  At the same time, the interaction of magnetic moments with conduction electrons in graphene is predicted to be different than what is observed in conventional metals,  offering new avenues to realize correlated electronic states\cite{shi2019kondo}.
 
 Here, we present an unexpected characteristic of defects in CVD graphene: we show that the relatively gentle  annealing process typically used to remove residues from graphene devices \cite{lin2011graphene,ishigami2007atomic,dan2009intrinsic,cheng2011toward,pirkle2011effect,dean2010boron,leong2014does,lin2012graphene} significantly enhances the quantum mechanical phase-breaking (dephasing) rate measured via weak localization (WL)\cite{tikhonenko2008weak}, and that this dephasing is due to the activation of magnetic moments. Although data from phase coherent measurements at cryogenic temperatures might seem unrelated to applications at room temperature, our results imply that these activated defects create a fast mechanism for spin relaxation that would limit the use of CVD graphene in spintronics.  At the same time, this finding sheds light on the microscopic origin of local moments in this form of graphene, and demonstrates that samples grown by CVD will be a fruitful platform to study interactions between magnetic moments and electrons in graphene.

\section{II. Experimental methods}
 Multiple samples were prepared by transferring commercial CVD graphene (ACS Materials Co., MA) onto n-doped Si wafers with a 295 nm SiO$_{x}$ dielectric.  In some samples, a 30 nm HfO$_{2}$ film was grown by atomic layer deposition (ALD) at \SI{100}{\celsius} on the SiO$_{x}$ before transferring graphene [Fig.~S1~inset].  
 Hall bars with a width and a length of 30 $\mu$m by 30 $\mu$m were defined by electron beam lithography.  Cr/Au (5/80nm) ohmic contacts were used to apply a bias current, $I$, and measure longitudinal and transverse voltages, $V_{xx}$ and $V_{xy}$ [Fig.~1(a)], then $V_{xx}$ was converted into resistivity $\rho$ or conductivity $\sigma\equiv\rho^{-1}$ using the geometric aspect ratio of the Hall bar.  The carrier density, $n_s$, was tuned using the backgate voltage, $V_G$, and measured via low-field Hall effect.
 
\begin{figure} [ht]
    \centering{\includegraphics[width=3.3in]{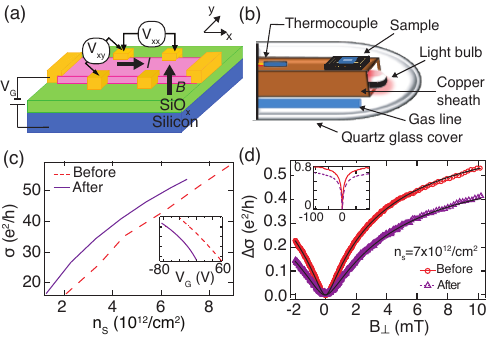}}
    \caption{
 (a) Schematic of the sample geometry and measurement setup. Pink indicates CVD graphene, yellow indicates Au/Cr electrodes. The dielectric stack shown here includes the SiO$_x$. (b) Schematic of the thermal annealing setup. (c) Change in conductivity as a function of charge carrier density, before and after Ar annealing. Inset: same data as a function of gate voltage. (d) Magnetoconductivity before and after Ar annealing, with fits to WL theory. Fit parameters before annealing: $\tau_{i}^{-1}$=160$\pm$45 ns$^{-1}$, \tp=33$\pm$2 ns$^{-1}$; after annealing: $\tau_{i}^{-1}$=215$\pm$90 ns$^{-1}$, \tp=60$\pm$5 ns$^{-1}$.  Fits included data over a larger field range \cite{supple}.
    }
    \label{fig1}
\end{figure}

Annealing was performed in Ar, N$_{2}$, forming gas (5\% H$_{2}$/95\% N$_{2}$, hereafter H$_2$/N$_2$), or under vacuum, using a simple thermal annealer [Fig.~1(b)].
Samples wire-bonded into ceramic chip carriers were loaded into the annealer, then pumped to a base pressure of $3\times10^{-3}$ mbar (gas anneal) or $<10^{-5}$ mbar (vacuum anneal). For gas annealing, the chamber was pumped-and-flushed, then the  flow rate adjusted using a needle valve until a pressure of 200 mbar was attained.   Samples were annealed for 1 hour at \SI{300}{\celsius}, then cooled to room temperature in the same atmosphere as the annealing step, and immediately transferred to the cryostat. After evacuating the cryostat and adding He exchange gas, samples were cooled to 4.2 K for measurement. The total air exposure time after annealing was less than one minute.  

\section{III. Result and discussion}
Figure 1(c) compares $\sigma(n_s)$ of a sample before and after annealing in Ar.  The conductivity itself rose slightly, while the gate voltage required to reach the charge neutrality point, $V_{n=0}$, decreased from 88 V to 18 V.
The quantum correction to the perpendicular magnetoconductivity, $\sigma(B_\bot)$, also changed after annealing [Fig.~1(d)]. To interpret this change, phase ($\tp$) and intervalley ($\tau^{-1}_{i}$) scattering rates  were extracted from $\Delta\sigma(B_\perp)\equiv\sigma(B_\perp) - \sigma(0)$ using fits to the standard WL expression for graphene~\cite{mccann2006weak, tikhonenko2008weak},
\begin{equation}
\Delta\sigma(B_\perp)
= \frac{e^2}{\pi \hbar}\left[F\left(\frac{\tau^{-1}_{B}}        {\tp}\right)-F\left(\frac{\tau^{-1}_{B}}    {\tp+2\tau^{-1}_{i}}\right)
\right],
\end{equation}

\noindent where $\tau_B^{-1}=4eDB_\perp/\hbar$, $F(z) = ln(z) + \psi (0.5+z^{-1})$, diffusion constant $D=\frac{\sigma \pi \hbar v_{f}}{2e^{2}\sqrt[]{\pi n_s}}$ and $v_f=10^6$ m/s for graphene.  A third term reflecting intravalley scattering is omitted here for clarity as it did not affect the extracted \tp\ or \ti, but was included in the fitting \cite{supple}.  

Both $\tau^{-1}_{i}$ and \tp\  rose due to annealing.  The rise in $\tau_i^{-1}$ reflects increased short-range scattering due to changes in the interaction between graphene and SiO$_x$ dielectric, caused by the annealing itself \cite{cheng2011toward}. The significant rise in \tp\ is more surprising, and is the central focus of this work.

Electron-electron (e-e) and electron-phonon (e-p) interactions are well known dephasing mechanisms in metallic systems. Below 50 K in graphene the e-e rate dominates, and at 4 K the e-p rate is negligible in comparison \cite{tikhonenko2008weak, tikhonenko2009transition, min2017asymmetric}.  The e-e dephasing rate in graphene is \cite{Altshuler_1982,aleiner1999interaction, mccann2006weak, tikhonenko2008weak, min2017asymmetric}:
\begin{align}
    \tee
 = \frac{k_{B}T}{\hbar}\left(\frac{ln\left(\frac{g}{2}\right)}{g}\right), ~g=\frac{\sigma h}{e^2}.
\end{align}

\begin{figure}[t]
    \centering{\includegraphics[width=3.3in]{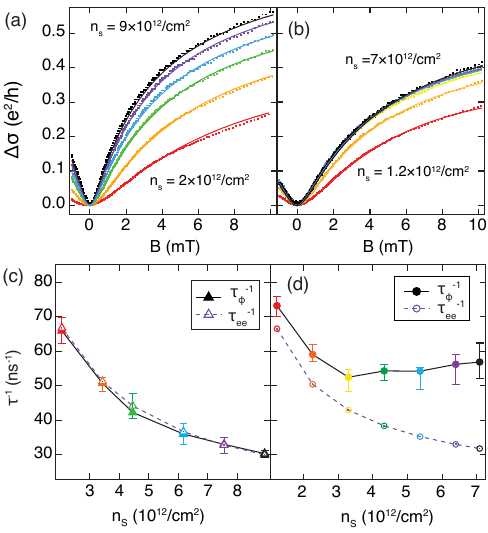}}
    \caption{
     The change of magnetoconductivity $\Delta\sigma(B)\equiv\sigma(B)-\sigma(B=0)$ with carrier density  before (a) and after (b) annealing in Ar.   Dephasing rates \tp, extracted from $\Delta\sigma(B)$ for each carrier density (filled/solid) compared with  e-e rates calculated using Eq.~2 (open/dashed), before (c) and after (d) annealing.  Error bars indicates range of possible values from fitting.
    }
    \label{fig2}
\end{figure}

For the data in Fig.~1(d), Eq.~2 predicts \tee= 33 ns$^{-1}$ before annealing, matching the measured value \tp=33$\pm$2 ns$^{-1}$.  After annealing, the calculated value from Eq.~2 is barely changed, \tee= 32 ns$^{-1}$, but the measured value rises to \tp=60$\pm$5 ns$^{-1}$.  Fig. 2 extends this analysis to a range of $n_s$.  Before annealing, measured \tp\ coincide almost exactly with calculated \tee\ [Fig.~2(c)]. After annealing, on the other hand, \tp\ exceeds \tee\ everywhere, with the difference, $\txs\equiv\tp-\tee$, growing larger with $n_s$ [Fig.~2(d)]. Taken together, Figs.~2(c) and 2(d) indicate that e-e interaction dominates dephasing in samples as deposited, but new interactions emerge in Ar-annealed samples that add 10's of ns$^{-1}$ of excess dephasing. 

\begin{figure}[t]
\centering{\includegraphics[width=3.3in]{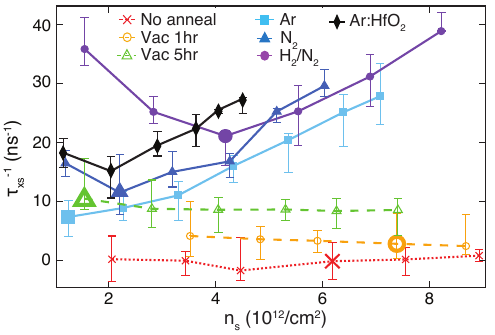}}
    \caption{
    The excess dephasing rate ($\txs\equiv\tp-\tee$)  as a function of  $n_s$, for various annealing conditions: no anneal; anneal in vacuum for 1 or 5 hours; and anneal 1 hour in Ar, N$_2$, or \fg. Longer anneals in gas environments were not investigated here.  Samples are directly on SiO$_x$ except where noted (Ar:HfO$_2$, black diamonds).  Larger symbols indicate the data point corresponding to $V_{G}$=0 for each recipe (for Ar:HfO$_2$, that point is off the graph and not shown). 
}
\label{fig4}
\end{figure}

As an inert gas, Ar is not expected to create chemical modifications to graphene at temperatures of only a few hundred $^\circ$C. Indeed, transmission electron microscope investigations have confirmed that anneals such as those performed here do not affect perfect graphene, but that polymer resist and other residues begin to break down in the 150 to 250 $^\circ$C temperature range, and radicals formed during that process may interact with dangling bonds at graphene defect sites \cite{lin2012graphene}.  Our observations are consistent with the reports of Ref. \onlinecite{lin2012graphene}, that the influence of annealing on  defect activation is via the thermal decomposition of residues rather than the gas itself.   

Figure 3 compiles the excess dephasing rate, $\txs$, for a variety of annealing recipes, and leads to several observations.  When annealing is performed in vacuum, \txs~is very small, rising only to $\sim$10 ns$^{-1}$ (constant in $n_s$) even after 5 hours of annealing.  For anneals in a gas environment, \txs~is significantly larger, growing with  $n_s$ by an amount that does not depend on the choice of gas.  \txs~also increases at low $n_s$, most noticeably for samples annealed in \fg.

Upon closer examination, the rise in \txs~at low $n_s$ in the \fg-annealed data from Fig.~\ref{fig4}a appears to be related to the gate voltage applied, rather than specifically to the reduction in carrier density, with the minimum in $\txs(V_G)$ consistently occurring at $V_G=0$. The charge neutrality point, $V_{n=0}$, was much larger for \fg ($V_{n=0}$=60 V) compared to Ar ($V_{n=0}$=18 V) or N$_2$ ($V_{n=0}$=31 V), so larger negative $V_G$ was applied to \fg-annealed samples to reach the low density regime.  Further insight into the importance of $V_G$, independent of $n_s$, is gained by measuring the $n_s$-dependent dephasing rate for a particular annealed sample, repeatedly shifting the Dirac point through exposure to ambient atmosphere  [Fig.~4] \cite{C0JM02922J}.  \ding{172} represents \tp\ for the sample immediately after an hour of annealing in \fg ($V_{n=0}$=60 V), \ding{173}-\ding{175} represent subsequent one hour exposures to ambient atmosphere ($V_{n=0}$=92,109,140 V), then \ding{176} represents an additional one hour anneal in \fg ($V_{n=0}$=70 V). 
Calculated \tee\  for the different exposures fall along a single curve [Fig.~4(a)], reflecting the fact that $\sigma(n_s)$ does not change through these different steps.   Measured \tp\ also fall along a single curve for the low-$n_s$ part of each scan, for \ding{172}-\ding{175}.  However, the onset of the upturn in \txs\ at high $n_s$ shifts to higher and higher densities with exposure to air, while the minimum in \txs\ remains pinned to $V_G=0$ [arrows in Fig.~4(b)].  This seems to indicate that the upturns in \tp\ and \txs\ at high density are connected  to the electric field between the graphene and the backgate rather than to the carrier density itself.

The persistent upturn in \txs\ at large $n_s$ (strongly negative $V_G$), even after extended exposure to air, indicates that the activation of a dephasing mechanism by annealing is robust.  A second annealing step (\ding{176}) following the multiple exposures to air results in $V_{n=0}$ shifting back to 70 V, while \txs\ increases above even the values observed after the first anneal (\ding{172}).  As was the case for \ding{172}-\ding{175}, the minimum in \txs remains at $V_G=0$.

\begin{figure}[ht]
    \centering{\includegraphics[width=3.3in]{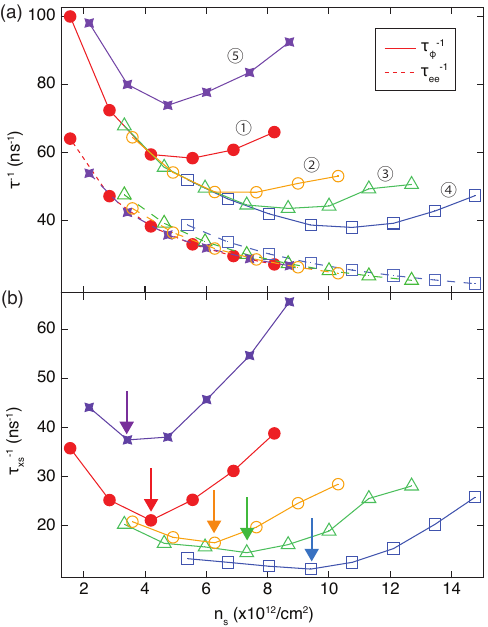}}
    \caption{
      (a) Evolution of the dephasing rate (solid lines) and e-e interaction rate (dashed lines), shown as a function of carrier density, for an \fg annealed sample after air exposure.  \ding{172} represents the sample immediately after one hour \fg anneal, \ding{173}-\ding{175} after consecutive air exposures, then \ding{176} after one more \fg anneal  (b) \txs calculated from the data in (a), with arrows indicating $V_{G}$=0.
} 
    \label{fig3}
\end{figure}

Just as the activated dephasing mechanism does not seem to depend on the choice of annealing gas, it also does not depend on the substrate.  In order to test for a possible influence from the substrate \cite{van2016spin, chen2012electronic,scarfe2021systematic,poljak2013influence}, Fig.~3  includes data from a sample, annealed in Ar gas, on a wafer where ALD-grown HfO$_{2}$ covers the SiO$_x$ [Fig.~S1~inset].  HfO$_{2}$ layers are amorphous when grown at the low temperatures used here, presumably with a defect density and type different from those found in thermal silicon oxide \cite{hausmann2003surface}. The mobility of the HfO$_{2}$ sample was similar to the samples on SiO$_x$, in contrast to previous reports of mobility enhancement by screening due to high-$\kappa$ dielectrics\cite{chen2009dielectric,konar2010effect}; this may indicate that the mobility in our samples is limited by defects in the graphene rather than charged-impurity scattering.  More importantly, the measured \txs\ for the HfO$_{2}$ sample is nearly identical to that of the SiO$_x$ sample \cite{supple}. From this, we tentatively conclude that the defects relevant to the dephasing enhancement lie in or on the graphene itself, rather than in the substrate.

The role of annealing temperature was also explored (data not shown).  Below \SI{200}{\celsius}, annealing had minimal effect on \txs\  for any of the inert gases.  At \SI{250}{\celsius} the effect was similar to, though less strong than, the \SI{300}{\celsius} data shown here.  For anneals at \SI{350}{\celsius} and above, the excess dephasing was even larger but the graphene mobility was  degraded.

\begin{figure} [t]
    \centering{\includegraphics[width=3.3in]{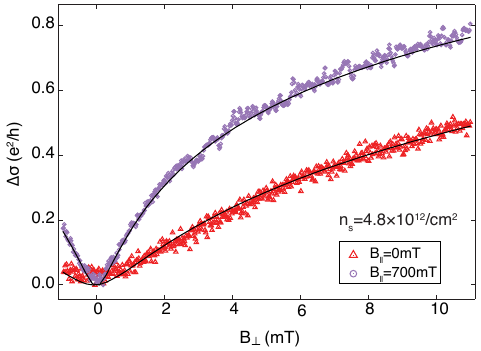}}
    \caption{Perpendicular field magnetoconductivity lineshape, $\Delta\sigma(B_\perp)$, for an H${_2}$N${_2}$-annealed sample changes dramatically with in-plane magnetic field, $B_\parallel$.  Solid lines are fits to Eq.~1. Extracted \tp  decreases from 95$\pm$10 ns$^{-1}$ at $B_{\parallel}$=0, to 24$\pm$3 ns$^{-1}$ at $B_{\parallel}$=700 mT (T=100 mK).}
    \label{fig5}
\end{figure}
\section{IV. dephasing mechanism}
 
The effect of an in-plane magnetic field on WL can reveal spin-related mechanisms for dephasing.  Magnetic fields, $B_\parallel$, applied in the plane of a graphene sheet influence $\sigma(B_\perp)$ in two ways.  First, ripples in the graphene create an out-of-plane local magnetic field, fluctuating randomly across the graphene even when  $B_\parallel$ is applied exactly along the average plane of the sample.  These out-of-plane components break time reversal symmetry locally, yielding an additional dephasing mechanism that grows quadratically with $B_\parallel$ \cite{lundeberg2010rippled}.  Second, $B_\parallel$ aligns local magnetic moments with Lande g-factor $g$ once $g\mu_B B_\parallel\gg k_B T$.  When free to rotate, these moments contribute to dephasing through spin flip scattering of conduction electrons, but that mechanism is suppressed when the moments are aligned by  $B_\parallel$ \cite{lundeberg2010rippled,lundeberg2013defect, kochan2014spin}.

Cooling the graphene samples in a dilution refrigerator equipped with a two-axis magnet enabled $\sigma(B_\perp)$ measurements with finite $B_\parallel$, at temperatures low enough that spins  could be aligned by $B_\parallel$ before dephasing due to ripples was significant. Fig.~5 shows WL data for an H$_2$/N$_2$-annealed sample at 100 mK. (This sample was annealed at \SI{350}{\celsius} to maximize \txs.)  
The lineshape of $\sigma(B_\perp)$  changes when in-plane field is applied, and fits to Eq.~1 confirm that \tp\ drops from 95 ns$^{-1}$  ($B_\parallel$=0) to 24 ns$^{-1}$ at  $B_\parallel$=0.7 T, where $g\mu_B B_\parallel/k_B T\sim 10$ with $g$=2.   Ripples would have induced the opposite effect (increased \tp for higher $B_\parallel$).  This data shows that spin flips induced by magnetic moments are the predominant mechanism for excess dephasing after annealing, contributing at least 70 ns$^{-1}$.

Magnetic moments are believed to emerge at graphene defects, including vacancies \cite{yazyev2007defect,chen2014magnetic},  impurities \cite{han2011spin,jozsa2009linear}, and strain fluctuations \cite{couto2014random,neumann2015raman}. In most cases, the interactions of defect-bound moments with conduction electrons are predicted to strengthen at low carrier density due to weaker screening ~\cite{chen2014magnetic,haase2011magnetic,yazyev2007defect,engels2014limitations}. We find, in contrast, that \txs\ increases at higher $n_s$ and, more importantly, depends more on the electric field due to $V_G$ than on the carrier density itself since gas annealed samples show the minimum near $V_G$=0. It may be, therefore, that the  magnetic moments being activated by gas annealing are of a different type than has been previously reported in experiment or studied theoretically.

\bibliographystyle{unsrt}
\bibliography{bibliography.bib}

\begin{thebibliography}{10}

\bibitem{lin2011graphene}
Yung-Chang Lin, Chun-Chieh Lu, Chao-Huei Yeh, Chuanhong Jin, Kazu Suenaga, and
  Po-Wen Chiu.
\newblock Graphene annealing: how clean can it be?
\newblock {\em Nano letters}, 12(1):414--419, 2011.

\bibitem{cheng2011toward}
Zengguang Cheng, Qiaoyu Zhou, Chenxuan Wang, Qiang Li, Chen Wang, and Ying
  Fang.
\newblock Toward intrinsic graphene surfaces: a systematic study on thermal
  annealing and wet-chemical treatment of sio2-supported graphene devices.
\newblock {\em Nano letters}, 11(2):767--771, 2011.

\bibitem{PhysRevB.22.5142}
B.~L. Altshuler, D.~Khmel'nitzkii, A.~I. Larkin, and P.~A. Lee.
\newblock Magnetoresistance and hall effect in a disordered two-dimensional
  electron gas.
\newblock {\em Phys. Rev. B}, 22:5142--5153, Dec 1980.

\bibitem{kozikov2010electron}
AA~Kozikov, AK~Savchenko, BN~Narozhny, and AV~Shytov.
\newblock Electron-electron interactions in the conductivity of graphene.
\newblock {\em Physical Review B}, 82(7):075424, 2010.

\bibitem{tikhonenko2008weak}
FV~Tikhonenko, DW~Horsell, RV~Gorbachev, and AK~Savchenko.
\newblock Weak localization in graphene flakes.
\newblock {\em Physical review letters}, 100(5):056802, 2008.

\bibitem{tikhonenko2009transition}
FV~Tikhonenko, AA~Kozikov, AK~Savchenko, and RV~Gorbachev.
\newblock Transition between electron localization and antilocalization in
  graphene.
\newblock {\em Physical Review Letters}, 103(22):226801, 2009.

\bibitem{min2017asymmetric}
Kil-Joon Min, Jaesung Park, Wan-Seop Kim, and Dong-Hun Chae.
\newblock Asymmetric electron-hole decoherence in ion-gated epitaxial graphene.
\newblock {\em Scientific reports}, 7(1):1--7, 2017.

\bibitem{neto2009electronic}
AH~Castro Neto, Francisco Guinea, Nuno~MR Peres, Kostya~S Novoselov, and
  Andre~K Geim.
\newblock The electronic properties of graphene.
\newblock {\em Reviews of modern physics}, 81(1):109, 2009.

\bibitem{geim2010rise}
Andre~K Geim and Konstantin~S Novoselov.
\newblock The rise of graphene.
\newblock In {\em Nanoscience and technology: a collection of reviews from
  nature journals}, pages 11--19. World Scientific, 2010.

\bibitem{RevModPhys.92.021003}
A.~Avsar, H.~Ochoa, F.~Guinea, B.~\"Ozyilmaz, B.~J. van Wees, and I.~J.
  Vera-Marun.
\newblock Colloquium: Spintronics in graphene and other two-dimensional
  materials.
\newblock {\em Rev. Mod. Phys.}, 92:021003, Jun 2020.

\bibitem{han2011spin}
Wei Han and Roland~K Kawakami.
\newblock Spin relaxation in single-layer and bilayer graphene.
\newblock {\em Physical review letters}, 107(4):047207, 2011.

\bibitem{lee2017review}
H~Cheun Lee, Wei-Wen Liu, Siang-Piao Chai, Abdul~Rahman Mohamed, Azizan Aziz,
  Cheng-Seong Khe, N~MS Hidayah, and U~Hashim.
\newblock Review of the synthesis, transfer, characterization and growth
  mechanisms of single and multilayer graphene.
\newblock {\em RSC advances}, 7(26):15644--15693, 2017.

\bibitem{yazyev2007defect}
Oleg~V Yazyev and Lothar Helm.
\newblock Defect-induced magnetism in graphene.
\newblock {\em Physical Review B}, 75(12):125408, 2007.

\bibitem{shi2019kondo}
Zheng Shi, Emilian~M Nica, and Ian Affleck.
\newblock Kondo effect due to a hydrogen impurity in graphene: a multichannel
  kondo problem with diverging hybridization.
\newblock {\em Physical Review B}, 100(12):125158, 2019.

\bibitem{ishigami2007atomic}
Masa Ishigami, JH~Chen, WG~Cullen, MS~Fuhrer, and ED~Williams.
\newblock Atomic structure of graphene on sio2.
\newblock {\em Nano letters}, 7(6):1643--1648, 2007.

\bibitem{dan2009intrinsic}
Yaping Dan, Ye~Lu, Nicholas~J Kybert, Zhengtang Luo, and AT~Charlie Johnson.
\newblock Intrinsic response of graphene vapor sensors.
\newblock {\em Nano letters}, 9(4):1472--1475, 2009.

\bibitem{pirkle2011effect}
A~Pirkle, J~Chan, A~Venugopal, D~Hinojos, CW~Magnuson, S~McDonnell, L~Colombo,
  EM~Vogel, RS~Ruoff, and RM~Wallace.
\newblock The effect of chemical residues on the physical and electrical
  properties of chemical vapor deposited graphene transferred to sio2.
\newblock {\em Applied Physics Letters}, 99(12):122108, 2011.

\bibitem{dean2010boron}
Cory~R Dean, Andrea~F Young, Inanc Meric, Chris Lee, Lei Wang, Sebastian
  Sorgenfrei, Kenji Watanabe, Takashi Taniguchi, Phillip Kim, Kenneth~L
  Shepard, et~al.
\newblock Boron nitride substrates for high-quality graphene electronics.
\newblock {\em Nature nanotechnology}, 5(10):722--726, 2010.

\bibitem{leong2014does}
Wei~Sun Leong, Chang~Tai Nai, and John~TL Thong.
\newblock What does annealing do to metal--graphene contacts?
\newblock {\em Nano letters}, 14(7):3840--3847, 2014.

\bibitem{lin2012graphene}
Yung-Chang Lin, Chun-Chieh Lu, Chao-Huei Yeh, Chuanhong Jin, Kazu Suenaga, and
  Po-Wen Chiu.
\newblock Graphene annealing: how clean can it be?
\newblock {\em Nano letters}, 12(1):414--419, 2012.

\bibitem{supple}
See supplemental material at [url will be inserted by publisher] for further
  details on the experiment.

\bibitem{mccann2006weak}
Edward McCann, K~Kechedzhi, Vladimir~I Fal’ko, H~Suzuura, T~Ando, and
  BL~Altshuler.
\newblock Weak-localization magnetoresistance and valley symmetry in graphene.
\newblock {\em Physical Review Letters}, 97(14):146805, 2006.

\bibitem{Altshuler_1982}
B~L Altshuler, A~G Aronov, and D~E Khmelnitsky.
\newblock Effects of electron-electron collisions with small energy transfers
  on quantum localisation.
\newblock {\em Journal of Physics C: Solid State Physics}, 15(36):7367--7386,
  dec 1982.

\bibitem{aleiner1999interaction}
IL~Aleiner, BL~Altshuler, and ME~Gershenson.
\newblock Interaction effects and phase relaxation in disordered systems.
\newblock {\em Waves in Random Media}, 9(2):201--240, 1999.

\bibitem{C0JM02922J}
Hongtao Liu, Yunqi Liu, and Daoben Zhu.
\newblock Chemical doping of graphene.
\newblock {\em J. Mater. Chem.}, 21:3335--3345, 2011.

\bibitem{van2016spin}
JJ~Van Den~Berg, Rositsa Yakimova, and BJ~Van~Wees.
\newblock Spin transport in epitaxial graphene on the c-terminated (000 1)-face
  of silicon carbide.
\newblock {\em Applied Physics Letters}, 109(1):012402, 2016.

\bibitem{chen2012electronic}
Kun Chen, Xiaomu Wang, Jian-Bin Xu, Lijia Pan, Xinran Wang, and Yi~Shi.
\newblock Electronic properties of graphene altered by substrate surface
  chemistry and externally applied electric field.
\newblock {\em The Journal of Physical Chemistry C}, 116(10):6259--6267, 2012.

\bibitem{scarfe2021systematic}
Samantha Scarfe, Wei Cui, Adina Luican-Mayer, and Jean-Michel M{\'e}nard.
\newblock Systematic thz study of the substrate effect in limiting the mobility
  of graphene.
\newblock {\em Scientific reports}, 11(1):1--9, 2021.

\bibitem{poljak2013influence}
M~Poljak, T~Suligoj, and KL~Wang.
\newblock Influence of substrate type and quality on carrier mobility in
  graphene nanoribbons.
\newblock {\em Journal of Applied Physics}, 114(5):053701, 2013.

\bibitem{hausmann2003surface}
Dennis~M Hausmann and Roy~G Gordon.
\newblock Surface morphology and crystallinity control in the atomic layer
  deposition (ald) of hafnium and zirconium oxide thin films.
\newblock {\em Journal of Crystal Growth}, 249(1-2):251--261, 2003.

\bibitem{chen2009dielectric}
Fang Chen, Jilin Xia, David~K Ferry, and Nongjian Tao.
\newblock Dielectric screening enhanced performance in graphene fet.
\newblock {\em Nano letters}, 9(7):2571--2574, 2009.

\bibitem{konar2010effect}
Aniruddha Konar, Tian Fang, and Debdeep Jena.
\newblock Effect of high-$\kappa$ gate dielectrics on charge transport in
  graphene-based field effect transistors.
\newblock {\em Physical Review B}, 82(11):115452, 2010.

\bibitem{lundeberg2010rippled}
Mark~B Lundeberg and Joshua~A Folk.
\newblock Rippled graphene in an in-plane magnetic field: effects of a random
  vector potential.
\newblock {\em Physical review letters}, 105(14):146804, 2010.

\bibitem{lundeberg2013defect}
Mark~B Lundeberg, Rui Yang, Julien Renard, and Joshua~A Folk.
\newblock Defect-mediated spin relaxation and dephasing in graphene.
\newblock {\em Physical review letters}, 110(15):156601, 2013.

\bibitem{kochan2014spin}
Denis Kochan, Martin Gmitra, and Jaroslav Fabian.
\newblock Spin relaxation mechanism in graphene: resonant scattering by
  magnetic impurities.
\newblock {\em Physical review letters}, 112(11):116602, 2014.

\bibitem{chen2014magnetic}
Jing-Jing Chen, Han-Chun Wu, Da-Peng Yu, and Zhi-Min Liao.
\newblock Magnetic moments in graphene with vacancies.
\newblock {\em Nanoscale}, 6(15):8814--8821, 2014.

\bibitem{jozsa2009linear}
C~J{\'o}zsa, T~Maassen, M~Popinciuc, PJ~Zomer, A~Veligura, HT~Jonkman, and
  BJ~Van~Wees.
\newblock Linear scaling between momentum and spin scattering in graphene.
\newblock {\em Physical Review B}, 80(24):241403, 2009.

\bibitem{couto2014random}
Nuno~JG Couto, Davide Costanzo, Stephan Engels, Dong-Keun Ki, Kenji Watanabe,
  Takashi Taniguchi, Christoph Stampfer, Francisco Guinea, and Alberto~F
  Morpurgo.
\newblock Random strain fluctuations as dominant disorder source for
  high-quality on-substrate graphene devices.
\newblock {\em Physical Review X}, 4(4):041019, 2014.

\bibitem{neumann2015raman}
Christoph Neumann, Sven Reichardt, Pedro Venezuela, Marc Dr{\"o}geler, Luca
  Banszerus, Michael Schmitz, Kenji Watanabe, Takashi Taniguchi, Francesco
  Mauri, Bernd Beschoten, et~al.
\newblock Raman spectroscopy as probe of nanometre-scale strain variations in
  graphene.
\newblock {\em Nature communications}, 6(1):1--7, 2015.

\bibitem{haase2011magnetic}
P~Haase, S~Fuchs, T~Pruschke, H~Ochoa, and F~Guinea.
\newblock Magnetic moments and kondo effect near vacancies and resonant
  scatterers in graphene.
\newblock {\em Physical Review B}, 83(24):241408, 2011.

\bibitem{engels2014limitations}
S~Engels, B~Terr{\'e}s, A~Epping, T~Khodkov, K~Watanabe, T~Taniguchi,
  B~Beschoten, and C~Stampfer.
\newblock Limitations to carrier mobility and phase-coherent transport in
  bilayer graphene.
\newblock {\em Physical review letters}, 113(12):126801, 2014.

\end{thebibliography}




\end{document}




\section{SUPPLEMENTARY INFORMATION}

\vspace{\baselineskip}

\textbf{Section 1. Thermal annealer}

\vspace{\baselineskip}

The sample rests on a copper sheath covering a halogen projector bulb.  The temperature of the copper sheath is monitored with a thermocouple, and the power of the bulb is adjusted with a variable transformer to keep the copper at the desired temperature. Sample temperatures were confirmed to match those of the copper using a mock sample with a fine-wire thermocouple glued on the wafer surface. Temperataures of the thermocouple and copper were found to match under both in gas and in vacuum annealing conditions.

\vspace{\baselineskip}
\vspace{\baselineskip}

\textbf{Section 2. Weak localization fitting process and analysis}

\vspace{\baselineskip}

Weak localization fits were made to resistivity data, instead of conductivity as expressed in Eq.~1 (main text), using 
$$
\rho(B) =\alpha(B-B_{0})+\rho_0-\rho_0^{2} \frac{e^2}{\pi \hbar} \left[ F \left( \frac{ \tau_{B}^{-1}}{ \tau_{ \phi }^{-1}} \right) -F \left( \frac{ \tau_{B}^{-1}}{ \tau_{ \phi }^{-1}+2 \tau_{i}^{-1}} \right) -2F \left( \frac{ \tau_{B}^{-1}}{ \tau_{ \phi }^{-1}+ \tau_{\ast}^{-1}} \right)  \right]
$$

\noindent The third term in the equation above, missing from Eq.~1, was found to be strongly suppressed by the enormous $\tau_*^{-1}$ in CVD graphene and therefore irrelevant to the extraction of \ti\ and \tp.  This equation adds several parameters to those two rates, which appear in Eq.~1:
\begin{itemize}
    \item An offset field $B_0$ was included to take into account the residual field in our superconducting magnet.
    \item The zero-field resistivity $\rho_0\equiv\rho(B=B_0)$.
    \item A linear-in-B term, $\alpha(B-B_0)$, was added to the resistivity, accounting for any offset of voltage probes that would mix Hall and longitudinal voltages.
\end{itemize}

Data were fit to this equation using a multi-step process involving two magnetic field ranges: $ \pm $10 mT [c.f. Fig.~1d] and $ \pm $100 mT [Fig.~S1].  In general, fine scans over the smaller field range enabled a more accurate determination of the smaller rate, \tp, whereas coarser data over the larger field range was important for determining \ti, and the value of \ti itself affected the determination of \tp.  The fitting process was:
\begin{enumerate}
    \item Determine $\rho_0$ and $B_0$ from Lorenzian fits to the peak of $\rho(B)$
    
    \item Fit $\pm$100 mT range to determine $\alpha,\tp,\ti$
    
    \item Fixing $\alpha, \rho_0, B_0$ and using \tp and \ti from above as initial parameters, fit $\pm$10 mT range.
    
    \item Fixing $\alpha, \rho_0, B_0, \tp$ and using \ti from above as initial parameter, fit $\pm$100 mT range.  If there is a significant change in \ti, repeat from step 2.
    
    \item Error bars for \tp were determined by trying different combinations of \ti and \tp to check how far \tp could be off before the fit was visually bad.
\end{enumerate}

\begin{figure}[ht]
	\centering
		\includegraphics[width=2.8in,height=2.3in]{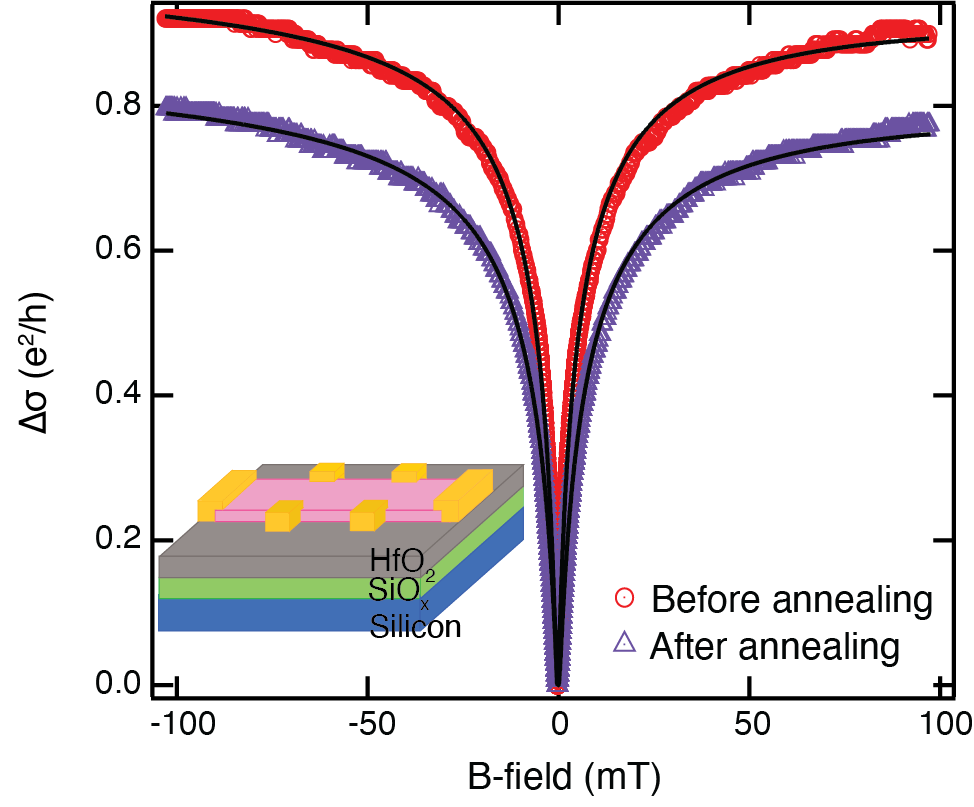}
	\caption{Magnetoconductivity before and after Ar annealing, with fits to WL theory (black solid lines), over the larger $\pm$100 mT range.  Inset: Schematic of the sample including a HfO$_2$ ALD layer (grey) on top of the SiO$_x$}
\end{figure}

\vspace{\baselineskip}

\textbf{Section 3. Graphene on HfO$_2$}

\vspace{\baselineskip}
Figure 3 in the main text shows \txs ~from a sample on HfO$_2$-covered SiO$_x$, annealed in Ar.  In addition to this data, a full set a measurements was taken on this sample.  Fig.~S2 shows the measured \tp ~and expected \tee ~before any annealing occured, confirming \tp~ is consistent with \tee.  After annealing in vacuum for 1 hour, \tp~ and \tee~ are nearly unchanged, again consistent with each other.  Only after annealing in Ar does \tp~ rise significantly above \tee, as it does for samples directly sitting on SiO$_x$.  One difference between graphene on HfO$_2$ and graphene on SiO$_x$ was that $V_{n=0}$ did not shift significantly due to the annealing on HfO$_2$, whereas on SiO$_x$ the shift was often several 10's of volts.  Instead, $V_{n=0}$ for the HfO$_2$ remained within 10 V of zero.

\vspace{\baselineskip}

\begin{figure}[ht]
	\centering
		\includegraphics[width=5in,height=2in]{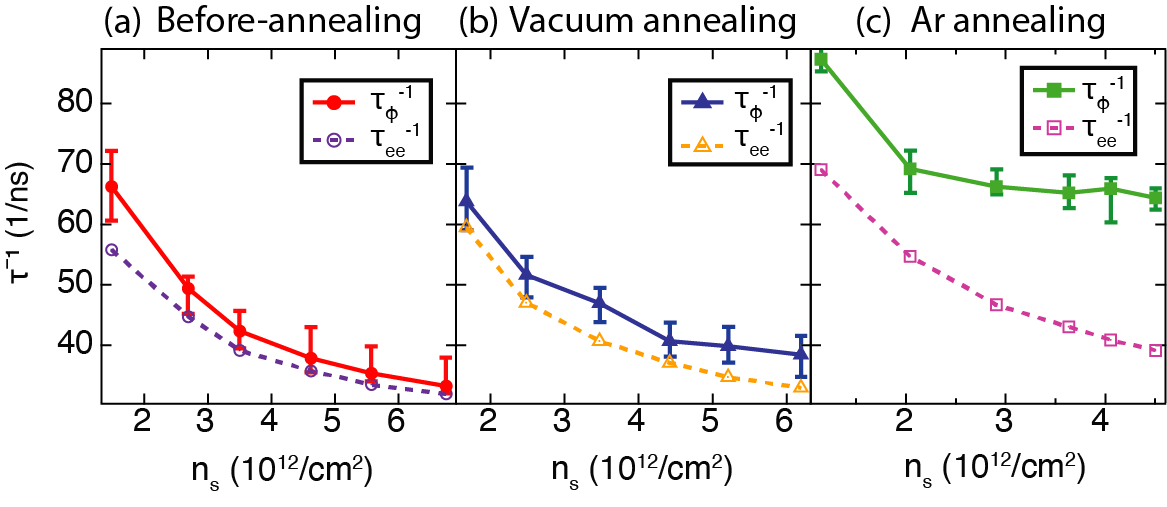}
	\caption{Measured (\tp) and expected (\tee) dephasing rates for graphene on a HfO\textsubscript{2}-covered substrate (a) before annealing, (b) after vacuum annealing, and (c) after Ar annealing. }
\end{figure}

\vspace{\baselineskip}

\vspace{2 in}






\vspace{\baselineskip}